\documentclass[acmsmall]{acmart}

\AtBeginDocument{%
  \providecommand\BibTeX{{%
    \normalfont B\kern-0.5em{\scshape i\kern-0.25em b}\kern-0.8em\TeX}}}

\setcopyright{rightsretained}
\acmJournal{PACMHCI}
\acmYear{2024} \acmVolume{8} \acmNumber{CSCW2} \acmArticle{502} \acmMonth{11}\acmDOI{10.1145/3687041}

\usepackage{subcaption}
\usepackage{multirow}
\usepackage{fontawesome}
\usepackage{makecell}

\begin{document}

\newcommand{\sys}{\textit{r/ifyoulikeblank}}

\newcommand{\papertitle}{Investigating Characteristics of Media Recommendation Solicitation in r/ifyoulikeblank}

\newcommand{\rqt}[2]{\hspace{5pt} \textbf{RQ#1}. \textit{#2}}
\newcommand{\rqsub}[2]{\hspace{10pt} \textbf{#1}. \textit{#2}}

\newcommand{\qt}[2]{\begin{quote}\small{\textit{``#2''}} \hspace{5pt} --- \textbf{#1}\end{quote} }

\newcommand{\addition}[1]{\textcolor{black}{#1}}

\newcommand{\inlineqt}[2]{\textit{``#2''}}

\newcommand{\out}[1]{{#1}}
\newcommand{\sang}[1]{\out{{\small\textcolor{blue}{\bf[*** Sang: #1]}}}}

\newcommand{\ssubsubsection}[1]{\subsubsection{\textbf{#1}}}

\newcommand{\rqone}{Why do people solicit recommendations in \sys?}

\newcommand{\rqtwo}{How do people seek recommendations in \sys?}

\newcommand{\rqthree}{How do people respond to and interact with recommendation requests in \sys?}

\title[\sys]{\papertitle}


% \settopmatter{authorsperrow=4}

% \author{Anonymous}
\author{Md Momen Bhuiyan}
\authornote{A part of this work was conducted while the author was at Virginia Tech}
\affiliation{%
  \institution{University of Minnesota Duluth}
  \city{Duluth}
  \state{MN}
  \country{USA}}
\email{mbhuiyan@d.umn.edu}

\author{Donghan Hu}
\affiliation{%
  \institution{Virginia Tech}
  \city{Blacksburg}
  \state{VA}
  \country{USA}}
\email{hudh0827@vt.edu}

\author{Andrew Jelson}
\affiliation{%
  \institution{Virginia Tech}
  \city{Blacksburg}
  \state{VA}
  \country{USA}}
\email{jelson9854@vt.edu}

\author{Tanushree Mitra}
\affiliation{%
  \institution{University of Washington}
  \city{Seattle}
  \state{WA}
  \country{USA}}
\email{tmitra@uw.edu}

\author{Sang Won Lee}
\affiliation{%
  \institution{Virginia Tech}
  \city{Blacksburg}
  \state{VA}
  \country{USA}}
\email{sangwonlee@vt.edu}

\renewcommand{\shortauthors}{Md Momen Bhuiyan et al.}

\begin{abstract}
Despite the existence of search-based recommender systems like Google, Netflix, and Spotify, online users sometimes may turn to crowdsourced recommendations in places like the \sys{} subreddit.
In this exploratory study, we probe why users go to \sys{}, how they look for recommendation, and how the subreddit users respond to recommendation requests.
To answer, we collected sample posts from \sys{} and analyzed them using a qualitative approach. 
Our analysis reveals that users come to this subreddit for various reasons, such as exhausting popular search systems, not knowing what or how to search for an item, and thinking crowd have better knowledge than search systems.
Examining users query and their description, we found novel information users provide during recommendation seeking using \sys. For example, sometimes they ask for artifacts recommendation based on the \textit{tools} used to create them. 
Or, sometimes indicating a recommendation seeker's \textit{time constraints} can help better suit recommendations to their needs.
Finally, recommendation responses and interactions revealed patterns of how requesters and responders refine queries and recommendations. 
Our work informs future intelligent recommender systems design.
\end{abstract}
% Recommender systems from Google, Netflix, and Spotify typically perform well recommending popular items while doing poorly for recommendations in users' personal long tails.
% new dimensions for consideration in a long-tail recommendation-seeking setting.

% rational behind a recommendation in \textit{Response}, expertise of a \textit{Recommender}, and  
% Or, description of the scene of the \textit{Event} where someone intends to use the recommendation is another consideration.




%    <concept>
%        <concept_id>10003120</concept_id>
%        <concept_desc>Human-centered computing</concept_desc>
%        <concept_significance>500</concept_significance>
%        </concept>
%    <concept>
%        <concept_id>10003120.10003121</concept_id>
%        <concept_desc>Human-centered computing~Human computer interaction (HCI)</concept_desc>
%        <concept_significance>500</concept_significance>
%        </concept>
%    <concept>
%        <concept_id>10003120.10003121.10003122.10003334</concept_id>
%        <concept_desc>Human-centered computing~User studies</concept_desc>
%        <concept_significance>300</concept_significance>
%        </concept>
%  </ccs2012>

% \ccsdesc[500]{Human-centered computing}
% \ccsdesc[500]{Human-centered computing~Human computer interaction (HCI)}
% \ccsdesc[300]{Human-centered computing~User studies}


\maketitle

\section{Introduction}
% Accuracy-based recommendation systems are widely deployed in the real world. However, over the years, many research indicated that focusing on accuracy could be insufficient since it tends to present users with popular items. Other characteristics such as novelty, unexpectedness and diversity should also be taken in to consideration.
% This notion becomes more applicable when it comes to suggesting recommendation to the users in their personal long tail.
% Though there are research that delves into how to incorporate such evaluation characteristics inside recommender systems, little is known about what users desire when it comes to searching for niche recommendation in their personal long-tail.
Recommendation seeking is a common activity for online users and they typically use a range of sources to do so, from using Google's algorithm to reading Amazon reviews from users.
Each of these methods has certain strengths.
For example, algorithm behind Google search may consider all the available information on the web with the history of the user when responding to search requests.
Whereas recommendation from a user can contain items the algorithms would not recommend.
More often than not, algorithmic systems provide very accurate results, especially when users are seeking popular content~\cite{Pelissari2022}.
However, these systems can sometimes produce insufficient results when users' needs are novel or diverse or very specific~\cite{cremonesi2010performance,konstan1997grouplens,Ho2014,mcnee2006being,tsai2018beyond}.
In such instances, some users may go for human-sourced or crowdsourced route through various platforms where they ask for recommendation from a stranger.
One particular place for such recommendation is \sys, a subreddit dedicated to ask for recommendations based on a set of examples and their latent similarities.
\addition{This example-based posting method allows recommendation seeking on  diverse types of media, unlike other subreddits dedicated on topics, such as \textit{r/movies} or \textit{r/music}.
Another important distinction is that, where dedicated subreddits spaces allows various forms of discussions, \sys{} allows recommendation related posts only.
Thus, it is a unique space for recommendation seeking online.
}
% However, accuracy-based recommendation systems perform well in terms of recommending popular items~\cite{Pelissari2022}, be it in solicited suggestions during a Google search or unsolicited suggestions while browsing the Netflix homepage (e.g., movies recommended under Comedy). 
% Here, when we speak of recommendations in ``long tails,'' we mean items in users' preferences that few other users prefer.
% This terminology is derived from the frequency distributions of users' artifact preferences, which typically behave like power law distributions with long tails---occupied by the least common preferences.
% Some past research has suggested adopting other evaluation criteria, such as novelty, unexpectedness, or diversity in recommender systems to address the inaccuracy issue in the long tail~\cite{herlocker2004evaluating}.
% Unexpectedness, for example, can lead to pleasant new findings for a user.
% However, such evaluation criteria still need to be considered within the context of users' personal interests in the long tail.
% That is, understanding the characteristics of recommendations users seek in their personalized long-tail still needed before considering these recommended evaluation criteria. 
% Learning the characteristics of recommendations users seek in their personalized long tails could help platform developers incorporate that information into context-aware machine learning models or help in the deployment of filtering options that users can use to refine results during recommendation-seeking activities.
Communities like \sys{} have been playing an important role in seeking and sharing recommendations for a long time, thereby affecting users' content consumption (note that \sys{} was created in 2011 but recommendation seeking from other users has a history as early as the history of the internet~\cite{konstan1997grouplens}).
With the rise of recommender systems over the years, user communities may now play a much smaller, more focused, and niche role. 
Today, people might go to such places for recommendations after exhausting other resources, such as general-purpose (Google) and specialized (Spotify) recommendation engines.
Despite \sys{} \addition{being among the top 1\% of subreddits} and being a space for recommendation seeking for about a million users \addition{for over a decade}, little is known about the characteristics of recommendation seeking here.
\addition{In the backdrop of the transformations happening in online recommendation seeking systems, from search systems (Google, Bing) to conversational recommender system (Bard, Copilot)~\cite{blogImportantNext,microsoftIntroducingMicrosoft}, it is crucial to understand how search systems were lacking and led people to go to subreddits like \sys.}
Therefore, in this exploratory content analysis, we sought to answer the following research question:

\begin{itemize}
    \item[\textbf{\textit{RQ1}}.] \textit{\rqone}
    \item[\textbf{\textit{RQ2}}.] \textit{\rqtwo}
    \item[\textbf{\textit{RQ3}}.] \textit{\rqthree}
\end{itemize}

% This is the goal of this research.
% Research in recommendation systems over the years have moved from item-similarity models to a context-aware models for some time.
% Various context has been considered during this time, such as, user persona, mood, intent,.
% To be more context aware through interaction, conversational recommender systems are also becoming more popular with the adoption of AI assistive smart home devices.
% However, an important question remain answered, \textit{what context matters to users and how conversational recommender should interact with their users?}
% Such context could be difficult to identify when users require recommendation in their personal long tail.
% % Though no framework for such context importance have emerged so far, especially in the personal long-tail.
% % motivation - informing future conversational agent interactions

% To fill this gap, we took a human-centered approach by conducting a content analysis of a popular subreddit called \sys{}, where users ask for recommendations from other users.

% \sys{} itself has its own culture and unique ways of enabling recommendation exchange activity.
% In their queries, recommendation seekers have to submit a list of items they find similar.
% They can explain what makes those items similar or leave it up to the commenters' to discern.
% Commenters can provide recommendation, with or without justifying them.
% Besides these constraints, there are very few rules of engagement in \sys.

% \rqt{}{}

% \rqt{2}{\rqtwo}

To answer our research question, we performed a qualitative analysis on recent posts and comments from this subreddit at the time of the study, which ran from January to September 2022.
During this period, our data set accumulated 19,715 posts and 119,610 comments.
Since answering our research questions needed analyzing posts, responses, and interactions, we sampled a set of \texttt{<Post, Comment, Original Poster's Reply>} triplets ($n = 1511$) from our data set.
(Note: Throughout this work, we will use ``original poster,'' ``OP,'' and ``poster'' interchangeably.)
% Among these triplets, 1000 came from music genre while the rest were from the combination of TV, Film, Game, and Books.
We applied a qualitative open-coding method to the sample to answer our research questions.
To complement the qualitative results, we also conducted some quantitative analysis.

For our first research question, we found that users come to this subreddit for various reason.
Some do after doing using both popular search methods (Google, Spotify) and specific tools  (\textsf{spotalike.com}, \textsf{rateyourmusic.com}).
Some do come when they are not clear on how to search for a specific thing or do not know which genre a content fall under.
Some also come because they think that users may have better expertise than search systems.
For our second, we found that the queries in \sys{} can be sorted into five high-level codes that contextualize various components of the recommendation-seeking setting
These components describe the parties involved items sought (\textit{Artist, Artifact}), its production (\textit{Production/Distribution}), context of the recommendation seeker (\textit{OP's Context}),  how request is formulated (\textit{Additional Information}).
% \textit{Artifact, Artist, Event, Non-primary Entity, Platform, Query, Response, Recommender, and Recommendation Seeker}.
% Except for Non-primary Entity and Platform, each code was further divided into several sub-codes.
For example, \textit{Artifact} contained such information as the message/theme inside it, the language it was created in, its creation time and location, its popularity among users, the genre it is in, and the tools used to create it.
Another code is \textit{Additional Information}, which describes such aspects as whether a query includes abbreviations of an artifact, whether a query relates to multiple types of artifact, whether a query includes images, whether a query contains justifications for choices, and whether it mentions a specific element inside an artifact.
We found approximately even number of mentions of artists and artifacts in post titles.
For our third research question, we found two high-level codes \textit{Response} and \textit{Interactions}.
In user \textit{Response}, we found subcodes such as user providing justification behind thier response, and user sharing previews of their suggestions to entice.
We found that \textit{Interactions} in the posts primarily consisted of subcodes, such as interaction affirming the taste of the seeker, long back and forth between two or multiple users, interaction endorsing other comments, interaction refining the original query, and recommendation seeker recommending in response to a recommendation from a commenter.
Quantitatively, we found that a large number of posts in \sys{} receive no responses.
Furthermore, various actions and interactions (e.g., the number of posts or comments by a user) in the community also follow a power law distribution.
Analyzing several search engines, including IMDB and Spotify, we found that a majority of our sub-codes are not supported as filtering mechanisms in these systems.
% A greater number of posts expressed emotions such as joy, love, optimism, and anticipation, while fewer posts expressed sadness, fear, and disgust.
% Emotional valence--wise, there were more positive posts than negative ones.
% Comparing the genres of music mentioned in \sys{} posts against several popular playlists from Spotify, we found that the genres of music in \sys{} posts were more diverse.
Overall, our results indicate opportunities for existing systems to adopt means of expanding users' expressivity in queries and diversifying the recommendation dimensions used in algorithms.
We conclude with a discussion on how to successfully seek recommendation from strangers, design dimensions for recommender systems and the challenges in incorporating our findings in existing recommender systems.
% , solicited and unsolicited recommendation, and conversational recommender system design simulating human-like interaction.

\section{Related Work}
In this section, we first provide background in the form of different types of recommender systems and features that have been considered for recommender system design in research.
Then, we discuss the issue of popularity bias and how that affects recommendation seeking.

\subsection{Recommender System Taxonomy and Feature Consideration}
Recommender systems are ubiquitous in various areas, such as shopping.
Much of the research conducted on recommender systems focuses on entertainment (music, movie, TV show, and book recommendations) and shopping~\cite{Katarya2016, Pera2014}.
These systems tend to take various criteria as input to provide recommendations, while the majority of the output of recommendation systems could be divided into either ranks or probabilities.
The earliest recommender systems (or search systems) primarily took content features, such as titles, descriptions, and related information as inputs.
From there, these systems would create an item-item similarity matrix; this method of recommendation generation is called content-based filtering~\cite{Vall2017}.
Since then, various other information has been incorporated into recommendation system designs.
Such information may include user information, such as demographics, personality traits (e.g., Big Five), habits, moods, and expertise~\cite{Lu2018,Andjelkovic2019,Lee2010}.
These additions resulted in the creation of metrics such as user-user similarity, known as collaborative filtering.
Some research has looked into context-based filtering by accounting for context, such as the time of the day, activities being performed by a user, and the presence or absence of a companion~\cite{Dias2014}.
Researchers also started incorporating user interactions into a system to improve recommendations.
Such interactions may include clicks, consumption duration, and textual input~\cite{Seifert2015}.
In some cases, eye tracking of the user has also been used~\cite{Gaspar2018}.
With the emergence of conversational agents, recommender systems can also take nuanced feedback, such as critiques, constraints, or restatements from users to improve recommendations~\cite{Cai2019, Cai2020, Lyu2021}.
Taken together, an a high-level perspectives, recommender systems can be divided into content-based filtering, collaborative filtering and hybrid approaches ~\cite{roy2022systematic}.
Furthermore, collaborative filtering can be further divided into memory and model based filtering.
In this work, we expand the domain and taxonomy of recommender system, particularly for the purpose of generating recommendations.
\addition{In the past, there has been some work exploring online user interaction to inform recommendation system design. For instance, Morris et al. investigated online question answering interactions on social media~\cite{morris2010people}. While there are some similarities, our investigation is focused on recommendation seeking, unlike the question-answering paradigm investigated by Morris et al.~\cite{morris2010people}.}

\subsection{Popularity Bias Issue during Recommendations Seeking}
When looking at peer-to-peer recommendations, it is important to be wary of the popularity bias problem---the tendency for popular items to be recommended frequently, decreasing accuracy and causing a lack of diversity in recommendations~\cite{Chen2014}.
In spite of this, few research works focused on improving other metrics, such as novelty and diversity, coverage, scalability, user trust, and satisfaction, which could reduce such bias~\cite{Pelissari2022}.
Although recommender systems can show many other biases, such as selection bias in data, position bias in produced recommendations, and bias amplification through feedback loops, this section focuses on popularity bias alone~\cite{Chen2022}.
Popularity bias has been identified and investigated for some time in recommender systems to understand the magnitude of its impact on these systems and their recommendation accuracy~\cite{Steck2011}.
These investigations have shown that popularity bias issue appears in different ways, mostly as content-agnostic factors, where users get popularity bias in both video and audio recommendations~\cite{Borghol2012, Yang2018, Aziz2022}.
When platforms ignore the popularity bias issue in recommendation results, that can cause problems, such as reduced personalization and fairness in recommendation results~\cite{Abdollahpouri2021}. 
This, in turn, result in poorer user experiences for users who prefer niche items, content that happens to be unpopular, and hidden gems.
Furthermore, training machine learning models on data (e.g., clicks) from popularity-based recommender systems can propagate the same problem in future systems~\cite{merton1968matthew,perc2014matthew}.

Due to this phenomenon, researchers have been finding solutions to mitigate or eliminate popularity bias. 
Optimizing for a different measure of the recommender system that the accuracy measure is a very common approach. 
Collaborative filtering and pairwise loss functions have utilized this approach for item popularity control and regularization, respectively~\cite{Chen2014,Lin2022,Rhee2022}. 
These methods have been studied to find the impact on the predictive behavior of the models after debiasing~\cite{Sun2019, Chen2022, Chen2021}. 
Another critical issue current research addressing popularity bias is the use of evaluation in an offline setting, often on a simulated dataset. 
Without examining them in an online setting some of the underlying problems might not get resolved~\cite{Yang2018}. 
Another attempted solution is to use a multi-armed bandit exploration-exploitation framework to diversify the recommender system results; this solution has produced good results in the discovery of new media~\cite{Parapar2021,Aziz2022}. 

Popularity bias can also be leveraged to improve recommender accuracy.
For example, some work has leveraged selection bias with user preferences to improve recommender system accuracy~\cite{Huang2022, Zhang2021a, Abdollahpouri2021}. 
This method can also be targeted for diversity or shifting consumption based other attributes, such as artifact age~\cite{Huang2022, Hansen2021}. 
% Using a user centered approach, we are able to optimize the recommender system's popularity bias to find new media that the user wants.

Additionally, these two approaches can be combined to create a system that eliminates or mitigates bias in a unique way. 
FAiR, a fairness-centric model, attempts such a method~\cite{Liu2022}.
Another method is to use a human in the loop to help with debiasing~\cite{Fu2021,bhuiyan2022othertube}.
The algorithmic technique above have a lacking focus on popularity bias from the perspective of data distribution which may contain imbalance due to differing audience preferences.
Understanding the audience preferences through the lens of recommendation-seeking processes in a crowdsourcing setting can complement the missing elements in typical search based recommendation seeking.
% We fill this gap through this work.

% \subsection{}

\section{Methodology}
To answer our research questions, we investigated the subreddit \sys{} by collecting recent sample data. Below, we present a summary of the subreddit with its rules and norms, followed by our data collection procedure.

\subsection{About \sys{}}
\sys{} is a community where users seek recommendations of ``music, television, video games, movies, or anything else'' from community members.
The subreddit was created on March 31, 2011 and has 907,000 members at the time of writing.
Creation of this subreddit can be linked back to a post in r/music~\footnote{\url{https://www.reddit.com/r/Music/comments/getg6/if\_you\_like\_you\_might\_like\_requests/}}, where a user asked about creating a thread for recommendation seeking based on a set of examples.
Subsequently, this subreddit was created as mentioned by one of the users in that r/music post.
Posts to \sys{} are subject to a set of guidelines, including these two:

\begin{itemize}
    \item Posts must be an ``If I like…'' [IIL] or ``If you like…'' [IYL] and must be formatted correctly.
    \item Be descriptive in your post and include at least one example in the title
    \item The maximum amount of examples per post is 9 and must include a text list
\end{itemize}

The top-level comments submitted under each post are typically answers provided by commenters.
In the ensuing discussion thread, there is no restriction on topics apart from a ban on ``political discourse or edgy joking''.
There are several ``flairs'' (categorization tags used on Reddit) available for tagging posts, including Music, Film, TV, Books, Games, Comics, Art, Podcast, and YouTube/Streaming.
Apart from these, there is another flair for self-promotion and playlists.

\begin{figure}[b]
  \begin{minipage}[t]{.55\linewidth}
    \centering
    \small        
    \captionof{table}{Summary statistics for the collected posts and comments}
    \label{tab:stat}
    \begin{tabular}{lr}
        \hline
        Total posts & 19715 \\
        Posts following formatting rules with at least one comment & 4183\\
        Total comments & 119610\\
        User comment--OP reply interaction pairs & 3834\\
        \hline
    \end{tabular}
    
  \end{minipage}\hfill
  \begin{minipage}[t]{.30\linewidth}
    \centering
    \small
    \captionof{table}{Samples for qualitative coding by flair}
    \label{tab:triplet}
    \begin{tabular}{rlrl}
        \midrule
        Music & 1000 & Games & 64\\
        Film & 185 & Image & 61\\
        TV & 167 & Books & 34\\
        \hline
    \end{tabular}
    
  \end{minipage}
\end{figure}

\subsection{Data Collection}
% To collect a meaningful sample of recent posts for analysis, 
We collected all the posts from January 1st, 2022, to September 20th, 2022, the time when the study began.
This sample accounted for recency, making the analysis results comparable to the state of modern recommender systems.
% posts from years back could contain attributes already supported by current recommendation systems.
We used the PSAW\footnote{\url{https://github.com/dmarx/psaw}} Python Pushshift API to collect the data.
This dataset contained 19,715 posts and 119,610 comments.
If we filter out posts whose formatting did not adhere to community rules, we are left with 4183 posts ($\approx 21\%$).
Out of our comment set, there are 3834 ($\approx 3\%$) interactions between original posters and commenters.
Table~\ref{tab:stat} summarizes this information.

\subsection{Sampling for Qualitative Coding and Coding Process}
Combining with the posts with commenter-OP comment pairs, we created 3834 triplets of <post, answer, OP's response>.
% The posts were tagged as the genre of Music, TV, Film, Games, Image, Misc, Podcast, documentary, Arts, Comics, and YouTube-Streaming. One of the authors recategorized the misc categories into the rest.
% Since this set was large, we sampled all the non-music comments and 1000 music.
% Furthermore, since Podcast (8), documentary (3), Arts (2), Comics (20), and YouTube-Streaming (4) had very few comments, we removed them from our sample.
Since these triplets skewed towards music content, we took all non-music items and sampled a 1000 music triplets for qualitative coding.
This process led to 1511 items from all the triplets, where 511 items were from non-music category and 1000 items from music category, creating a better balance for analysis.
Note that we categorized the items primarily according to assigned post flairs, coding items independently only when posts had not been assigned flairs.
Table~\ref{tab:triplet} shows this distribution.

We started an open-coding annotation process for the selected question--answer--OP response triplets sampled from \sys~\cite{khandkar2009open}.
% To be more specific, there might be multiple answers and OP responses under one question. OP and other Reddit users could start discussions under one response for further information or even start a new topic.
Initially, we split the sample evenly between authors; each author went through their portion and performed coding.
In the coding process, each author made a record of anything they noticed not covered by previous codes.
For example, they could write down a new statement describing what they noticed and include an example: ``The author looks for songs with symbolism or hidden meaning''. 
From this phase, each author provided a list of codes which we later combined into a final list.
After this phase, we had the first version of our codebook, consisting of a set of 168 statements.
This list included some duplicates across content categories since authors identified similar codes across categories. 
Then, one of the authors iterated over these statements to combine similar codes into one and create a high-level taxonomy with codes and sub-codes.
After that, all the authors finalized the taxonomy through a discussion.
Table~\ref{tab:rq1} shows the final scheme.
In this table, to further contextualize how the characterisitics described there compare against the available search options in popular search systems, we explored the advanced search characteristics on four search systems (IMDB, Spotify, Steam, and Amazon Books), relating to the four types of media recommendation users seek in \sys, i.e., Movies/TV, Music, Games, and Books. 

% Table \ref{tab:rq1} shows the compiled codebook.

\section{RQ1: \rqone}
Our analysis of the sampled dataset revealed various characteristics relating to our first research question about why people come to \sys.
Below, we discuss them.

\subsection{Inadequacy in Other Methods}
Many of the description in \sys{} query indicated that users come to the subreddit after they have tried other methods.
For example, some mentioned they have tried popular methods like googling or trying spotify search. However, these systems returned results that did not match the criteria users had (\inlineqt{U18}{I googled and all the suggestions are based on the plot, but idgaf about the plot}). Sometimes, the results were harder to parse for the recommender seeker (\inlineqt{U20}{I did googled similar shows but there are so many of them and I don't know witch one would be worth a time}). 
% \qt{U19}{i googled this song and learned that the artist is jpop, but that is certainly not what im looking for as i dont even enjoy the vocals and prefer this instrumental version}
% \qt{U21}{The band has more listeners on Spotify but they are still a bit obscure, so I haven't found any good recommendations when googling}
% \qt{U7}{I’ve tried googling this but I’m mostly being recommended other thematically similar shows (like lost) instead of shows with a similar genre or vibe. What are some good high stress mystery drama type shows?}
% \qt{U9}{The beat drop and lyrics are amazing. I need suggestions on songs like this one!! I’ve tried listening to songs that Spotify tells me are similar to it but they just don’t have the same feeling. I need the great minds of all you redditors to help me}
% \qt{U14}{I've tried searching on Goodreads but nothing I've found has interested me.}
% \subsection{After Trying Specific Tools}
Some of the users also mentioned specific tools they have tried besides common search tools.
They mentioned several music searching options like spotalike.com and rateyourmusic.com.
Here, Spotalike is a search tool that returns spotify playlists similar to any track user searches for. Rateyourmusic is another music discovery tool.
% \qt{U4}{I've tried tools like Spotalike, but haven't found any songs that capture the same feeling.}
% \qt{U5}{I tried looking up the genres on rateyourmusic, but then you get people like Bladee, who is kind of bad and cringe imo (ecco2k is the only person like that that's kind of good, drain gang has like one good song)}
% \subsection{After Searching in Specific Genre}
% \subsection{After searching in other posts in \sys{}}
% \qt{U17}{I tried different methods (checking all jazz albums, tapping and singing it to the song-finding apps, doing research on Google)}

% \subsection{Out of Desperation}
A few users mentioned they had searched for some time and resorted to the subreddit afterwards as a last method to try (\inlineqt{U11}{I have tried for months to find anything similar and it's just not happening}). Overall, this theme indicate users' desperate attempt to find similar content after exhausting other methods.

% \qt{U1}{I tried searching for something similar but wasn't satisfied. I doubt I'll have any luck here but thought I should try ... }
% \qt{U8}{Looking for a specific sound in dubstep,"ive been in love with fox stevensons older music (2013-2015) for quite a while and i cant get my finger on what i can look for to find more music like it. specifically something similar to his songs ""out there"" and ""like you"". ive tried searching things like ""high energy dubstep"" but it has failed me.  if anybody has any suggestions PLEASE let me know!!!}

% \qt{U16}{I'm sure someone else has asked this on here, I've searched all the posts I can find relating to this, and I've tried listening to all the reccomendations.}

\subsection{When It Is Unclear How to Search for Content}
A few users mentioned they were looking for certain content but was not clear how to search for them.
Some were unsure whether there was a name for the genre of content they were looking for (\inlineqt{U13}{Not sure if there is a name for this style.}).
In some instances, users even asked for keyword recommendation to identify such content (\inlineqt{U22}{I really like the art style in the role playing game Mörk Borg, of which I know almost nothing. I'd like to find some terms that help me with googling more of the same style, and/or artists that do this kind of stuff.}
). 
% \qt{U3}{I'm really not even sure if it has a name to it. I tried looking up ""dark ambience"", but none of the songs have fit. It's like ambience, but gloomy, but also has some action or life to it? I can't describe it but I'm not a musican, so if anyone could help me identify the genere of these pieces, I'd be really grateful}
% \qt{U6}{Does anyone know what i'm thinking of? i've tried searching for all key words that could connect to the video but can't find it. Might be on Vimeo, thats what i remember}
Some users also mentioned going through various genre of music to find related items but failing to find what they were looking for (\inlineqt{U10}{I tried to search through a lot of genres and subgenres, such as Hardcore, J-Core, Gabber, Cybergrind, Darksynth, etc.}).
Now, these cases indicate content genre that are either new to users or harder to express in search systems (\inlineqt{U2}{I tried to find some other songs that make me produce such mass amounts of dopamine in the electro swing genre...}).

% \qt{U15}{I've also tried searching for other ""gypsy jazz"" type of music, and there's some good songs out there, but again, none that get me as good as this song.}

\subsection{Thinking \sys{} users Having Better Expertise than Search Systems}
In a few cases, users mentioned coming to \sys{} because they thought that users in the subreddit were more experts in finding the content they liked.

\qt{U12}{The problem is that Spotify is no longer capable of helping finding similar songs. I've tried for hours, but I guess that plenty of people here has good suggestions :)}

\noindent Overall, these three methods indicated primarily why users came to \sys{} for content recommendation, elaborated in discussion.

\section{RQ2: \rqtwo}
% Below, we present the result of our analysis outlining the characteristics of queries, responses, and interactions developed through our coding process, along with a summary of our quantitative exploration of \sys{}.
% Our coding process led to the development of a codebook with 10 codes and 37 sub-codes in total.
In our analysis, we found five high-level aspects or codes relate to the characteristics of questions and answers.
Notice that, due to nature of qualitative analysis,this result is not perfect; it provides only a naïve representation.
We discuss these codes in detail below.

\begin{table}
    \centering
    % \footnotesize
    \scriptsize
    \caption{Codes and sub-codes identified in qualitative open coding, answering our research question. Here, the Flair column shows the categories in which we encountered a given subcode. Some of these characteristics have been mentioned in prior literature. This reference list is not exhaustive. In addition to the breakdown of the result, we investigated four existing platforms whether they support for each sub-code as a filtering option or not, and included the result from there in this table. Here, * indicates limited support.}
    \label{tab:rq1}
    \begin{tabular}{rlllcccc}
        & & & & \multicolumn{4}{c}{\textbf{Platform Supporting the Feature}}\\
        \cline{5-8}
        & & & & & & & \textbf{Amazon}\\
        \textbf{Code} & \textbf{Sub-code} & \textbf{Flair} & \textbf{Prior Refs.} & \textbf{IMDB} & \textbf{Spotify} & \textbf{Steam} & \textbf{Books}\\
        \midrule
         \multirow{9}{*}{Artifact}  &  Message/Theme  &  Games, Music  & \cite{Lyu2021, cremonesi2010, bountouridis2019} & \faicon{check} & \faicon{check} & \faicon{check} & \faicon{check}  \\ % Cuisine Type/Menu in a restaurant
         &  Adjectives  &  Music, TV  & & \faicon{check} & \faicon{check} & \faicon{check} & \faicon{check}\\
         &  Popularity  &  Music  & \cite{Lyu2021,Seifert2015, Sedhain2013} & \faicon{check} & \faicon{check} & \faicon{check} & \faicon{check} \\ % rating
          &  Date  &  Games, Music  & &  \faicon{check} & \faicon{check} & \faicon{check} & \faicon{check}\\
          &  Genre  &  Books, Film, Games, Music, TV  & \cite{Dias2014,Lyu2021, dhakad2017} & \faicon{check} & \faicon{check} & \faicon{check} & NA \\ % content feature to find weights, place with scenic view/Ambience/Noise level
          &  Language  &  Music, TV  & \cite{Seifert2015} & \faicon{check} & \faicon{check} & NA & \faicon{check}  \\ % for search results: preferred language of results
          &  Location  &  Film, TV  & \cite{Lyu2021, lu2016} & \faicon{check}* & NA & NA & NA  \\ % restaurant location
          &  Tools  &  Music  & &  NA & NA & NA & NA \\
        \hline
        \multirowcell{4}{Production/\\ Distribution}  &  Completeness  &  TV  & & NA & NA & NA & NA   \\
          &  Revision  &  Music  & &  NA & NA & NA & NA \\
          & Non-primary Entity  &  Games, Music  & & \faicon{check}* & \faicon{check}* & \faicon{check}* & \faicon{check}*\\
          & Platform  &  Games, Music, TV  & & NA & NA & \faicon{check}* & \faicon{check}* \\
        \hline
        \multirow{4}{*}{Artist} &  Genre  &  Music  & & \faicon{check}* & \faicon{check}* & \faicon{check}*  & \faicon{check}* \\
         . &  Activity Level  &  Music  & & NA & NA & NA & NA  \\
          &  Demography  &  Music  & \cite{dai2014} & \faicon{check}* & NA & NA & NA \\
          
          &  Popularity  &  Music  & &  \faicon{check}* & \faicon{check} & NA & \faicon{check}  \\
        \hline
        \multirow{7}{*}{OP's Context}  &  Activity  &  Books, Film, Games, Music, TV  & \cite{Dias2014,Lyu2021,Zheng2016, mcinnis2016} & NA & \faicon{check} & \faicon{check} & NA \\ % activities were walking/relaxing/running/sleeping/shopping, enjoyable with family, family
          &  Scene  &  Games, Music  & \cite{Zheng2016} & NA & \faicon{check} & NA & NA\\ % time/day/season/location/weather
           & Expertise/Acknowledgement  &  Games, Music  & \cite{Seifert2015} & NA & NA & NA & \faicon{check}*\\ % level of expertise in a information searching context, books have reader age
          &  Feeling  &  Games, Music  & \cite{Andjelkovic2019,Zheng2016} & NA & \faicon{check} & \faicon{check} & NA \\ % mood is tagged for the song by the artist, emotion
          &  Idiosyncrasy  &  Music  & & NA & NA & NA & NA \\
          &  Time Commitment  &  Film, Games  &  & NA & NA & NA & NA \\
          &  Willingness to Repeat  &  Film, Games  & & NA & NA & NA & NA\\
        \hline
        \multirowcell{5}{Additional\\Information}  &  Abbreviation  &  Film, Games, TV  & & \faicon{check} & \faicon{check} & \faicon{check} & \faicon{check}\\
          &  Transcendence  &  Books, Games, TV  &  & NA & NA & NA & NA\\
          &  Image  &  Books, Music  & \cite{feldman2020} & NA & NA & NA & NA\\
          &  Justification  &  Games, Music  & \cite{Tsai2020,Lyu2021} & NA & NA & NA & NA\\ % users were asked to provide reason they choose an item, justify
          &  Specificity  &  Books, Games, Music, TV  & & NA & NA & NA & NA\\
          & Desired Recommender  &  Music  & \cite{Lee2010,Lee2014} & NA & NA & NA & NA\\ % experts are those who listen to novice users recommendation in the short tail
        \hline
        \hline
        \multirow{3}{*}{Response}  &  Justification  &  Film, Games, Music  & \cite{Cai2020,Lyu2021} & NA & NA & NA & NA\\
          &  Preview  &  Film  & \cite{jadidinejad2019} & NA & NA & NA & NA\\
          &  Transcendence  &  Music  & & NA & \faicon{check}* & NA & NA\\
        \hline
        \multirow{7}{*}{Interaction}  &  Affirming Taste  &  Music  & &  NA & NA & NA & NA \\
          & Discussion  &  Books  & \cite{Lyu2021} & NA & NA & NA & NA\\ % personal opinion, Comparison, Persuasion, Prior Experience
          & Endorsement  &  TV  & & NA & NA & NA & NA\\
          & Refinement  &  Books, Film, Games, Music, TV  & \cite{Cai2020,Lyu2021} & NA & NA & NA & NA\\
          & Repetition  &  Books, Games, Music  &  & NA & NA & NA & NA\\
          & Reciprocity  &  Music  & & NA & NA & NA & NA\\
        & Miscellaneous  &  Books, Film, Games, Music, TV  & \cite{Cai2020,Lyu2021,Seifert2015} & NA & NA & NA & NA\\ %,,recommendations desirable?
        \hline
    \end{tabular}
    % \vspace{-8pt}
\end{table}

% \sang{reviewed up to this point. }

\subsection{Query - Artifact}
% The code ``Artifact'' describes the thing user is asking recommendation for, i.e., music/song, movie, tv series, books etc. Since this is the primary reason for posting on \sys{}, it is not surprising that a large number of codes (n) described the artifact. We found these description within the question, with a few in the answer and op's response. Within the artifact code, we found several subcodes describing various aspects of the artifact.
In this context, an Artifact is an object of the class within which a user is seeking recommendations, such as music/songs, movies, TV series, and books, among others. \sys{} rules ask users to mention up to nine Artifacts in post titles. We will look closer at their presence in our quantitative analysis. 
% Given the fundamental purpose of the posting here is seeking such artifacts, it is not surprising that we found nine subcodes used to describe the artifact. 
% These codes were identified within the question, with a few instances observed in the answer and the original poster's response. 
Within the Artifact code, nine sub-codes were employed to capture various facets of a given artifact. We discuss them here:

\ssubsubsection{Message/Theme}
% Content describes what the artifact consists of. 
% For example, content of a song is what the lyrics says or refers to or content of a movie is the story it describes. 
% When asking about content, OP did primarily seek certain things (e.g., biblical reference in a song) are there or whether any story exists inside (e.g., a series of albums tells a complete story).
% In games, OP asked for content with certain musics, dialogs, and stories.
% Though our dataset only referred to content for music and games, this subcode can also be extended to rest of of the artifacts.
Message/Theme refers to the substance of an artifact, which encompasses various elements central to its creation and interpretation. For example, the message of a song includes the lyrics and the themes that it expresses or refers to, while the content of a movie is the story that it tells, along with its characters, settings, and messages. Users seek specific details or elements that can help them better understand and appreciate an artifact when inquiring about content. In our data set, original posters (OP) sought content, such as biblical references in a song or the presence of a coherent story throughout a series of albums.

In the case of games, message includes such elements as musical scores, dialogues, and storylines which contribute to the overall gameplay experience. For example, ``looking for horror themed stories about private investigation'' appeared in a post in which the OP sought content that encompassed specific types of musical score, dialogue, and storyline to enhance their gaming experience. It is worth noting that while we found this subcode within content types of music and games, the subcode can also apply to other types of artifacts. 
Furthermore, users could be open to recommendations for multiple types of artifacts sharing such content attributes.
% With regular expression, we found 557/19715 instances of this subcode (e.g., ``old folk song in the story'' and ``looking for horror themed stories about private investigation'').
This is one of the common filters available across all the search systems we examined.
However, availability does not necessarily mean the recommendations from these systems match users' expectations.

\ssubsubsection{Adjective}
In a few instances, OPs applied certain adjectives (surreal, mind-bending, cathartic) to a particular class of artifact.
These adjectives could sometimes be categorized as content genres (mind-bending), but not always (surreal, steamy).
Similar to abbreviations, genre-like adjectives are supported by recommender platforms in our list.

\ssubsubsection{Popularity}
For music, we found some OPs were interested in items with \textit{less} popularity.
One reason behind this requirement could be that they wanted to explore more music that they had not experienced before. 
While more popular artifacts are easily discoverable using online recommender platforms, the same cannot be said for less popular artifacts.
% Conducting regular expression check, we found 1287/19715 instances of mention of popularity.
However, this number includes the popularity of both artifacts and artists.

\ssubsubsection{Date}
OPs and responders both mentioned Artifacts coming from certain time periods.
We found OPs mentioning specific date ranges or names in titles and posts to clarify their requests.
Examples of such requests include music from a specific time frame (2011-2017), TV series with content depicting a certain kind of period (``futuristic dystopia''), and game backgrounds from certain times (``middle Ages'' and ``the 50s'').
Some of these items may intersect with Genre subcode. 
% Using Spacy Entity detection, we found 626/19715 mention of time among the posts.
All four recommender systems from our list already support searching artifacts by these characteristics, either in metadata through filters or keyword searches.

\ssubsubsection{Genre}
A large number of observations were related to the genre of an artifact.
This code varied by content category.
In the context of music, OPs requested temporal characteristics (tempo, length of a song), atmospheric and storytelling qualities (triumph, epic, confident), and musical genres (fusing multiple genres).
In the context of games, they requested artifacts with a specific perspective for the main character (first vs. third person).
In the context of books, OPs requested protagonist or other character attributes (LGBT, outstandingly smart), high-level theme (fantasy), and meta-characteristics (stories about stories).
For TV and Film, OPs requested storytelling styles (non-linearity), dressing styles (costume, cool outfit, cowboy, leather), and high-level theme (mind-bending, psychological, rom-com).
% We found 5943/19715 reference to style in our dataset.
Generally, existing recommender systems allow artifact searches using this characteristic, and some unsolicited recommenders (e.g., Netflix home) often use this characteristic (e.g., rom-com) to categorize their recommendations.

\ssubsubsection{Language}
In the TV Series and Music categories, OPs commonly asked for artifacts in a certain language.
These searches included asking for translated versions of artifacts in other languages, such as with subtitles or voice-over translation.
For example, an English-speaking user might be happy to be recommended anime in Japanese as long as English subtitles are available.
In other instances, such as a documentary or reaction video, the OP may instead request commentary or narration in a particular language.
Similar to other sub-codes, this one is also extensible to many types of artifacts, such as movies.
% For example, someone might be interested in movies in a certain language.
% We found 742/19715 mentions of languages in our dataset.
Among the four recommender systems we checked, only IMDB, Spotify, and Amazon Books allow filtering by language.

\ssubsubsection{Location}
Location was a requirement for some OPs, who sought artifacts with geographic requirements at different levels. 
For example, OPs included region-level requirements (shows that take the time to incorporate elements of a given town or city) or landscape-level requirements (``movies that take place in the rainforest'').
We found that some posters specified examples to clarify their needs (e.g., ``I’m not searching genre-based, instead I’m searching for shows that make me feel like I’m living in those places. Californication in Cali, Justified [in] Kentucky, The Americans [in] DC'').
One OP mentioned that the rationale for this requirement was that they wanted artifacts that made them feel like they were in a certain place.
% In our post set, we found 457/19715 instances of location using SpaCy entity detection method.
Since this relates more to movies or tv series, IMDB seems to allow this filtering.
However, their filtering is cruder than the examples we mentioned here.

\ssubsubsection{Tools}
Sometimes, OPs requested artifacts created using specific tools.
For instance, in music, OPs asked for music featuring particular instruments, such as guitar, piano, and violin.
For TV and Films, they requested particular cinematography tools.
They also asked for certain combinations of instruments.
% Using a list of dictionary of musical tools\footnote{\url{https://github.com/imsky/wordlists/blob/master/nouns/music_instruments.txt}}, we found 1575/19715 matches to this characteristic in our dataset.
Among the four search platforms, none support filtering by this characteristic.

\subsection{Query - Production/Distribution}
In addition to artifact characteristics, queries in \sys{} sometimes also mentioned characteristics relevant to the elements of production and distribution of an artifact, outlined below.

\ssubsubsection{Completeness}
Completeness indicates whether the creation or publication of an item is finished or ongoing. 
For instance, a user in our data set requested recommendations for complete TV series. 
Completeness could be described in various terms, including regarding a particular or final season. 
This category also applies to other artifacts, such as musical albums, movie series, book series, and so on. 
In some instances, we noticed that users preferred incomplete artifacts they could consume incrementally during the release period.
% Using linguistic analysis with regular expression matching, we found 67/19715 instances of this subcode being mentioned in our post sample.
None of the four platforms we checked support this feature as an option for filters or recommendation criteria.

\ssubsubsection{Revision}
Primarily in the context of music, some recommendation seekers sought out artifacts that were revisions of original artifacts.
These revisions had several typologies, such as remixes, covers, and even stylistic similarity.
Sometimes, an OP would ask for new content in the style of older content (e.g., ``a modern band creating music like Nirvana'').
Outside of music, this code could apply to adaptations of books into movies or TV shows.
% We looked keywords related to this characteristics and found 191/19715 matches.
Although these characteristics are not supported as a filter on the platforms we checked, some cues in artifact titles indicating revision status (e.g., ``cover by'', ``[remix]'') can be discovered in regular keyword searches. 

% \qt{OP}{I'm looking for a good anime with LGBT protagonists, queer themed or at least with significant queer storylines.}

\ssubsubsection{Non-primary Production Entity}

Besides asking for recommendations based on an artifact or artist, there were a few instances where the OP asked for recommendations based on entities other than the primary artist involved in the production of an artifact. These requests included such entities as record labels for music, producers of movies, and the publisher of a game.
We did not identify any sub-codes under this code.
Among our four recommendation platforms, all of them provided some support for this feature as a filter.
For example, IMDB allows searching by company, and Spotify allows searching by label.

\ssubsubsection{Platform}
Since many artifacts are only available on certain online platforms, some posts also included a platform criterion.
For music, OPs might ask for songs available on Spotify.
For games, they might look for content playable on certain devices (mobile phone, PC, Xbox).
For TV shows, they might seek out content on Netflix (Netflix Originals) or YouTube.
This code also applies to regional availability, such as, Netflix India may have some content not available in Netflix USA.
% Our regular expression matcher identified 226/19715 matches to these keywords among the posts.
Among the four platforms, both Steam and Amazon Books provide partial sets of options to filter recommendation results by platform.
For example, Steam has filters for platform tags like Xbox, PC, Mac, and others.

\subsection{Query - Artist}
Next, we found many queries either described or asked for Artists---the primary creators of artifacts. We separated supporting production agents into a separate category. Although posters asked for Artists only in the context of music, the sub-code could apply to other categories, such as Movies. For Artists, we found four sub-codes: Artists' demography, theme of work, popularity, and activity level.
Below, we elaborate on them.

\ssubsubsection{Genre}
Similar to the genre aspect under Artifact, several posts characterized an artist's genre as a requirement.
Such requirements included styling factors determined by the artist, such as the structure of content created, patterns of music (melancholic or slow-burn), popular categories (indie, Latin, jazz, rock), layering (building up various components), performance style of sections within a song (yelling or breaking down).
% Using regular expression, we found 1682/19715 mention of this characteristic in our post datset.
Within our set of platforms, we found limited support for this through keyword searches and genre filters.

\ssubsubsection{Activity Level}
For music, several data points indicated that OPs were interested in artists who were active creators at the time.
These activities were related to aspects such as music production and whether they were on tour (e.g., ``I’m trying to find similar stuff [to the band The Clash, specified in the title] by newer bands that are still making music and possibly touring. Ideas?''
This attribute is not supported as a filter in our platform set.)
%, while the next subcode, i.e., demography could act as a supplement for this subcode to a certain extent. 

\ssubsubsection{Demography}
In several posts, OPs provided Artists' demographic backgrounds as criteria. Background information specified included artists' ethnicities (people of color), races (Black or White), individual or group gender identities (e.g., ``I m looking for something with screamy female vocal''), and relationships between artists within a group (e.g., siblings).
% Performing regular expression matching, we found 1003/19715 posts contained this subcode .
Among our list of platforms, IMDB has limited support for this, including searching by age and gender.

\ssubsubsection{Popularity}
In addition to activity level, some posters were interested in finding Artists that were less well-known or famous.
In one instance, OP asked for songs from less-known bands.
Therefore, this sub-code seems to be similar to its counterpart in Artifact.
With the exception of Steam, the other three platforms in our list seem to support this attribute as a filter or category (e.g., ``Hidden Gems'').

\begin{table}
    \centering
    \small
    \caption{Patterns. Punctuation marks and ``and'' separate items.}
    \label{tab:pattern}
    \begin{tabular}{p{0.45\linewidth}p{0.10\linewidth}p{0.40\linewidth}}
        \textbf{Pattern} & \textbf{\# Match} & \textbf{Matching Examples}\\
        \hline
        <song name> by <artist name> & 4721 & Click by Charlie, Blackstar by David Bowie, world.execute(me) by Mili \\
        these <song, artist, music> & 648 & these artists, this guy, These songs  \\
        <music, song, band> <like or about> <string> & 384 & artists like Girl Talk, Songs about suicide\\
        <song, music> that <verb> like <string> & 41 & song that sounds like it's the new single off your favorite band's best album \\
        \hline
        \multirow{5}{*}{Items matching no pattern} & \multirow{5}{*}{5648} & Rush E (Fanchen's rendition) as study music\\
        & & Genre and Record Label Compilation Albums\\
        & & Narrated music that tells a story\\
        & & Hip hop songs with beats like this?\\
        & & songs with dreamy instrumentals and fast lyrics\\
        \hline
    \end{tabular}
    % \vspace{-8pt}
\end{table}

\begin{table}
    \centering
    \small
    \caption{Strings to relevant item detection using google search }
    \label{tab:song_dist}
    \begin{tabular}{llll}
        \textbf{Type} & \textbf{\# among Matched Patterns} & \textbf{\# among Leftover} & \textbf{\# among All (unique)}\\
        \hline
         Artist & 203 & 1862 & 2065 (1292) \\
         Track & 1332 & 910 & 2242 (2029) \\
         Album & 149 & 223 & 372 (334) \\
         \hline
         All & & & 4679 \\
         % Tracks & & & 2029 \\
         \hline
    \end{tabular}
    
    % \vspace{-8pt}
\end{table}

% \ssubsubsection{Artists vs Artifacts}
So far, we have seen artists and artifacts are two primary aspects mentioned in \sys{} queries.
Between these two, one may ask, \textit{what do users mention more in post titles?}
To answer, we examined how often users mention between these two in music posts.
Particularly, we looked for artists, tracks, and albums in post titles of posts receiving responses.
% since the rest of the data was harder to clean.
For this purpose, we first split strings on punctuation and ``and'', which resulted in 11442 strings.
Next, we manually examined the text and identified four patterns.
These patterns were: <song name> by <artist name>, these <song, artist, music>, <music, song, band> <like or about> <string>, <song, music> that <verb> like <string>.
For example, ``Blackstar by David Bowie'' matches the pattern <song name> by <artist name>.
We used the Spacy NLP toolkit~\footnote{\url{spacy.io}} to identify the strings that matched these patterns.
Table~\ref{tab:pattern} shows counts of these patterns with example matches, as well as examples of strings not belonging to any pattern.
After this, we used Google searches to identify whether the patterned strings referenced an artist, track, or album.
We used Google search because it enabled us to identify items better, even for strings not matching any pattern in the previous steps.
This process involved searching Google with each string and checking if the search results included a link to Spotify or Apple Music.
This process led to identification of 4679 artists, tracks, and albums.
Table~\ref{tab:song_dist} shows the resulting distribution.
Here, we found that artists and tracks were mentioned pretty evenly and held the majority, with relatively few (8\%) mentions of albums.
% We also noticed that 40\% of the artist references were duplicates.
Overall, artist and artifact mentions in titles indicate even interests by \sys{} users in each category.

\subsection{Query - OP's Context}
To elaborate on why users seek recommendation, we found various information relating to the context an OP was in (activity they were doing) and any other constraint they may have relating to their search (time commitment). Below, we outline these subcodes.

\ssubsubsection{Activity}
Some posters asked for artifacts to play during an activity.
Such activities included physical activities, such as workouts, dance and party,. Some wanted to create playlists for these activities.
Some also wanted to play (i.e., cover) particular songs themselves.
Typically, these requests also include targets the posters intend to involve: a multiplayer game to play with friends/family, a movie to watch with their significant other.
OPs may even mention the number of people involved (``4 player split screen game'').
% Sometimes, OPs would ask for content that could act as a memorabilia, for instance, for their dog.
Both Spotify and Steam support searching by this characteristic (e.g., ``workout music'').

\ssubsubsection{Scene}
In several cases, OPs requested recommendations to play in various scenarios.
Such scene descriptions included locations (restaurants and cafes), other artifacts (songs or background music for a movie or commercial), abstract items (paintings), and environment (COVID isolation).
Using linguistic analysis, we found 3025/19,715 references to this attribute in our post set.
Among the four platforms, only Spotify has options for finding items by this attribute.

\ssubsubsection{Expertise/Acknowledgement}
In their queries, posters sometimes mentioned that they were uncertain about the ideal terms for what they were seeking. In such cases, they provided elaborate descriptions.
This strategy indicated that these users had a novice level of expertise.
Conversely, when OPs responded to comments, we noticed that some of them were receptive to trying lists of recommended items whenever they recognized at least a few items within the list (see quote below).
None of the platforms in our list are able to provide comparable information, indicating the extent of their knowledge in recommended results or specifying an expertise level. %; however, Amazon Books support a proxy feature through the ``Reader Age'' option.

\qt{OP}{[I] Like some King Tuff. Black Moon Spell is a banger. will definitely try out your other suggestions too.}

\ssubsubsection{Feeling}
In the contexts of music and games, OPs mentioned their current state of mind as a way to find matching artifacts.
Such criteria included emotional states (sad, happy), vibes (getting over an ex, dark, drug addiction), or emotional responses to content consumption (eerie).
For example, one OP wrote ``I feel so safe whenever I watch these. They're so comforting.''
We have outlined the varying emotion and sentiment distributions of the posts in our list in our quantitative analysis section.
This characteristic is supported by three out of the four platforms in our list.
For example, Spotify allows listeners to find music by mood.
However, these options are likely limited to popular genres of music. 

\begin{table}
    \centering
    \small
    \caption{Emotion and sentiment of posts.}
    \label{tab:emotion}
    \begin{tabular}{cc}
        \begin{tabular}{rlrlrl}
        \hline
        Joy          & 7903 & Sadness & 2136 & Surprise  & 29 \\
        Optimism     & 5359 & Disgust & 329  & Pessimism & 27 \\
        Anticipation & 3667 & Anger   & 308  & Trust     & 2  \\
        Love         & 2894 & Fear    & 287  &           &    \\ 
        \hline
        \end{tabular}
        &
        \begin{tabular}{rl}
            \hline
            Positive & 5071\\
            Neutral & 4752\\
            Negative & 519\\
            No majority & 9373\\
            \hline
        \end{tabular}
    \end{tabular}
    % \vspace{-8pt}
\end{table}

To further contextualize this subcode, we quantitatively looked at emotions and sentiments present in all the posts (both title and body) using TweetNLP, a library primarily devised for tweets which also works well for short texts like Reddit posts~\cite{tweetnlp}.
Each of the models used in the classifiers was a fine-tuned version of roberta\footnote{\url{https://huggingface.co/cardiffnlp/twitter-roberta-base-2021-124m}} on TweetEval\footnote{\url{https://huggingface.co/datasets/tweet_eval}} tasks.
The result of this analysis is shown in Figure~\ref{tab:emotion}.
Note that the emotion classifier could assign multiple labels to a single post, so the sum can exceed the number of posts.
We found positive emotions such as Joy, Optimism, Anticipation, and Love ran high in the majority of the posts.
Still, 15\% of the posts showed negative emotions such as Sadness, Disgust, Anger, and Fear.
Looking into the emotional valence, we found that in half of the posts, no particular valence (positive, neutral, or negative) was dominant.
However, among the rest, a majority (95\%) were positive or neutral, with the remaining 5\% expressing negative sentiments.

\ssubsubsection{Idiosyncrasy}
In rare cases, OP mentioned certain idiosyncrasies about who they are compared to what they like.
For example, one OP said, ``So I'm well aware my music taste is that of a 14-year-old TikTok obsessed, mentally ill Pinterest kid, and trust me I'm trying to get out of it.''
This sub-code could be covered by other characteristics for artifacts or artists.
However, we created this as a separate sub-code with a view towards capturing some of the nuances of the recommendation seekers.
This characteristic is not supported by the four platforms in our list.

\ssubsubsection{Time Commitment}
One particular characteristic that affects recommendations is the availability of a user to consume the content.
Some requesters are able and willing to devote a large amount of time to consuming recommended content, such as playing a 100-hour game or binge-watching a 100-episode TV series.
Requesters expect to receive recommendations that fit these contraints.
The platforms on our list do not support this characteristic.

\ssubsubsection{Willingness to Repeat}

Another characteristic that came up in several posts was whether posters would be open to repeating (rewatching or replaying) things they have consumed before.
In the words of one OP, ``Oh, maannnn!!! I haven't watched Ryan Trecarten's films in a while. I appreciate the reminder and the list.''
% Analyzing the data, we found 226/19715 matches about this openness to repeat.
Recommender systems can take this habit of recommendation seekers into consideration (e.g., Netflix's Watch it Again).

% To sum, these were the 9 high-level codes and underlying subcodes we found relevant to this research question.

\subsection{Query - Additional Information}
Examining the posts, we found that there were some aspects in query that did not fall under any codes described so far.
We identified several characteristics here, including the use of abbreviations, adjectives, transcendence, images, justification, and specificity, described below. 

\ssubsubsection{Abbreviation}
Posts sometimes used abbreviations for certain kinds of content.
Some of the abbreviations were ones that enjoy broad use (HIMYM for ``How I Met Your Mother'', DnD for ``Dungeons and Dragons'').
In some cases, posters created ad hoc abbreviations for their posts and reused them to refer to an item throughout the post text.
Abbreviations are generally recognized by the platforms in our list, although it can depend on how well-known the acronyms are. 

\ssubsubsection{Transcendence}
When seeking recommendations, OPs sometimes asked for recommendations that transcend multiple categories.
We termed this Transcendence (of the query scope).
For example, an OP might provide references in one category, like music, and ask for recommendations in other categories, like movies or TV shows (``lo-fi hip hop. Are there movies/TV shows that capture that vibe?'').
This characteristic is not supported by any platforms in our list.

% \qt{OP}{Video games, movies, shows, anything accepted but preferably (preferably) no books and podcasts (unless exceptionally good :-) ).}
% while being surrounded by advanced robotic animals. This new world has their own worries too, with these problems both big and small acting as the focus of both games.

\begin{figure}
    \centering
    \includegraphics[width=0.7\textwidth]{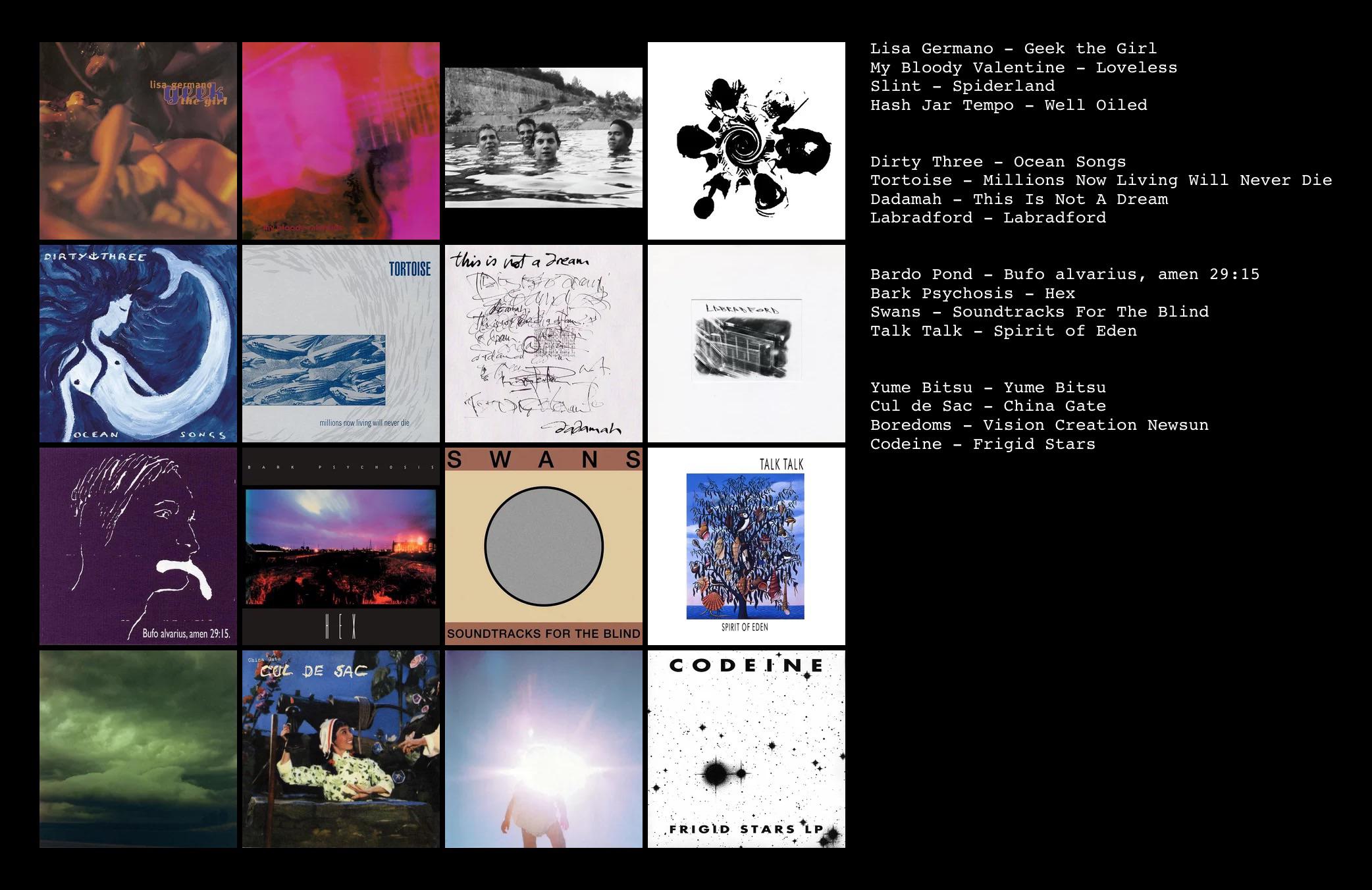}
    \caption{Grid of images for an example query.}
    % \vspace{-8pt}
    \label{fig:grid_image}
\end{figure}

\ssubsubsection{Image}
Some posters also made requests by posting sets of images of albums, songs, or books.
The purpose of formatting queries this way seemed to be to include particular art patterns not describable sufficiently through texts.
See Figure~\ref{fig:grid_image} for an example.
This input format is not supported by any platforms in our list.

\ssubsubsection{Specificity}
In several queries about different artifacts (music, games, and books), OPs provided different types of specific information in their queries.
These details were provided in terms of starting or ending time points (``0:32 second'', ``opening'', ``first 30 seconds'') for music and particular styles of writing within an author's works (``ending of under the dome by Stephen King'') for books.
Sometimes, these details were broad or nuanced (``Relaxing games that I can play with no stress'').
The four recommender systems in our list currently do not provide support for this feature.

\ssubsubsection{Desired Recommender}
In several instances, we found that one aspect of recommender came into attention, i.e., their expertise.
For example, a recommender may not be familiar with all the items in a request while nonetheless having suggestions to offer.
The OP in one such instance was still receptive to the suggestions despite the knowledge gap.
This varying degree of expertise is not typically realized by current recommender systems, let alone by those on our list.

\qt{Recommender}{I actually haven't heard of any of the artists you've listed (I will have to check them out), but as for Canadian hip-hop artists, are you familiar with Cadence Weapon? I believe he won a Polaris prize a few years back.}

\begin{figure}
    \centering
    \includegraphics[width=0.75\textwidth]{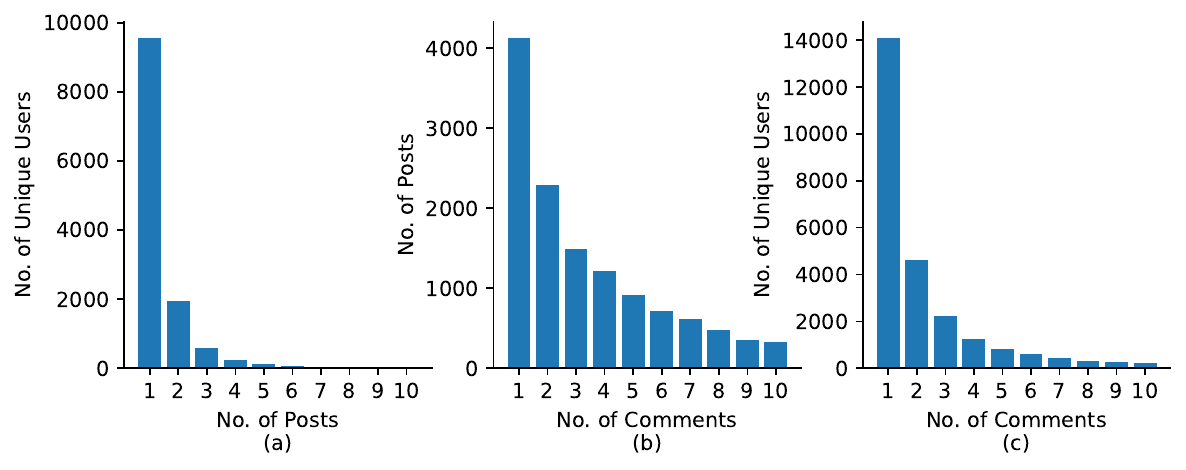}
    % \vspace{-8pt}
    \caption{Distribution of user participation in \sys. (a) Y No. of user submitting X number of posts (b) Y No. of posts having X number of comments (c) Y No. of user submitting X number of comments. X axis is capped at 10. }
    \label{fig:dist}
\end{figure}

\section{RQ3:\rqthree}
In response to this research question, we initially quantitatively looked at user participation in \sys.
Figure~\ref{fig:dist} shows user participation in \sys{}, showcasing the distribution of the posts and comments by users.
While this subreddit Figure~\ref{fig:dist}(a) shows that most posters (75\%) on \sys{} are one-time posters, with only 15\% posting twice.
Similarly to posters, a majority of commenters (57\%) on \sys{} are also one-time responders, with only 19\% commenting twice, as shown in Figure~\ref{fig:dist}(c).
Looking into comment count by posts that receive at least one comment, in Figure~\ref{fig:dist}(b), we found that 77\% of those posts received up to five comments.
Overall, we see that the interactions on this subreddit also look like a power law distribution with a long tail.
Our qualitative analysis further revealed two high-level codes relating to this research question, discussed below.

\subsection{Response}
In our analysis, certain characteristics were prevalent in the recommendations provided by the commenters in their response. We found three such characteristics, discussed below.

\ssubsubsection{Justification}
When responding, commenters sometimes provided justification; that is, the reason they thought their recommendations matched the OP's criteria.
Sometimes, justifications were descriptive in nature; other times, they consisted of ratings, either from commenters themselves or from external reviewing sites.
Posters themselves sometimes responded to explain why a commenter's recommendation did not match their criteria.
See the examples of these interactions below.
Among our set of platforms, none provides justification. However, other platforms like Netflix provide limited justification (e.g., ``Because you liked [X]'').

% \qt{Recommender}{Bloodborne. Although, the Lovecraftian elements[OP's requirement] don't really show up until about halfway to two-thirds of the way through the game.}

\qt{Recommender}{Hollow Knight (Darkest game on this list, but easily the cutest and most beautiful. Give it time as the opening hours can sometimes put off new players. 10/10)}

% \qt{Recommender}{The Other Guys: \url{https://www.rottentomatoes.com/m/other_guys}}

\qt{OP}{Nightwish [a recommendation provided by the commenter], obviously talented vocalists (Tarja and Floor both), but symphonic power metal is not my cup of tea.}

\ssubsubsection{Preview}
In some cases, commenters provided responses that included previews or snippets to entice the OP.
These previews could be taken from a trailer or certain section of an artifact (``The funeral scene [\url{https://www.youtube.com/watch?v=Oyj7YeXHhcA&ab_channel=Angie}] is amazing.'').
This feature is not supported by the platforms on our list.

\ssubsubsection{Transcendence}
In their answers, recommenders sometimes provided content outside of the requested category, such as a game instead of a book (see quote below).
% Some would list out single items while others may lists such as playlists.
This is a counterpart to the similar characteristics under Query.
Among our list of platforms, Spotify provides a similarly generalized response through its search function, which can return matching tracks, playlists, albums, or artists.

\qt{Recommender}{[In response to book recommendation] Do stories from video games count? If so, the Horizon games, Zero Dawn and the just released Forbidden West, would interest you. The Earth of the past was destroyed, but humanity was preserved and started to rebuild ... }

\subsection{Interaction}
Besides the characteristics of the response, our qualitative analysis revealed several themes prevalent in user interaction. 
These characteristics are not supported by the four platforms we examined.
However, these sub-codes could be incorporated inside conversational recommender systems to provide a better user experience, i.e., inside LLM-based conversational agents like ChatGPT or Google's Bard.
% We discuss them below.

\ssubsubsection{Affirming Taste}
In their interactions, commenters would sometimes respond just to express appreciation for OP's taste without offering a recommendation (e.g., ``great taste in music! just came to say that'').
In some cases, they would reveal that they had tastes similar to those of the OP (``We're like twins [with similar taste in music]'').
This type of interaction not only validates the OP's taste but also has the potential to create bonds between strangers on \sys.
Conversational recommender systems could promote a better human-like impression with this feature.

\ssubsubsection{Endorsement}
Among the interactions, in some instances, one commenter would endorse another's response with a comment (e.g., ``[In response to some other commenter's recommendation] Strongly seconding Alma's Not Normal and This Country - both are really great recs based on OP's list.'').
This does not necessarily add further information.
Although Reddit has upvotes and downvotes to operationalize endorsement, this behavior is still seen often.

\ssubsubsection{Refinement}
As mentioned previously, we found 12 long interactions focused on refining recommendations.
Besides these, there were also shorter interactions where OP and commenters would try to refine criteria.
In some cases, the OP would provide exclusion criteria from the beginning (e.g., ``Nothing against death metal btw, just not what I'm looking for here.'')
OPs sometimes edited their posts to add criteria  (e.g., ``edit: similar vibe and love story stuff overall something not over corny like a notebook but not zero love and overall corniness little more or little less ...'').
Also, OPs could add criteria in response to a commenter's suggestion after identifying a mismatch (e.g., ``[in a comment] Although I really enjoy these tracks, it's not what I'm looking for exactly. It's tough to explain, it's the way the lead synth comes together with the bass which gives it a certain feel ...'').
In some instances, OPs mentioned that some of the suggestions from commenters were not available on their preferred platforms (e.g., ``she sounds great :) such a shame that I can't find her on Spotify'').
Sometimes, commenters sought clarification from OPs about vague or unclear requirements (e.g., ``What do you mean by `sometimes actively disliked' the institute?'').
% \qt{Commenter}{Do the suggestions need to have romance / a love story, or just similar vibes?}
OPs could also ask for clarification from commenters when recommendations were not specific enough---for instance, when there are multiple movies with the same name (e.g., ``Thanks! Also which year was The Guilty released? There's about 5 movies on IMDB with that name, and is the Carnage one the 2011 one?'').
To simulate these refinement interactions, a recommender system needs to hypothesize the gap first and implement some reasoning capacity.
% When commenters are not sure whether an item fits OP's requirement, they would express that in their comments.
% \qt{Commenter}{Majesty was[sic] somewhat close}

\ssubsubsection{Repetition}
Sometimes, an OP would request the same recommendations over and over by posting a question repeatedly.
For example, one OP said, ``I'm sure I've asked similar questions to this in the past but I'm honestly always looking for more of this type of music''.
A recommender system needs to account for this type of interaction through memory and refine its responses appropriately.

\ssubsubsection{Reciprocity}
When commenters responded by appreciating an OP's taste or indicating in any other way that they had similar tastes, the OP would sometimes provide some recommendation back to those commenters as a form of reciprocity.
This quote from an OP illustrates one such instance: ``Also, if you're interested, check out 'men in grey' by Hellfreaks. More hard rock than anything, but vocally you might like the vibe''.

% \qt{OP}{So far, we're up the same ally. Thank you for these suggestions. Takyon's great... have you heard the \href{https://soundcloud.com/annoyingringtone/annoying-ringtone-takyon-def-qon}{Annoying Ringtone dancecore flip} of it?}

\ssubsubsection{Miscellaneous}
Many interactions between users fell into this category.
For instance, OPs responding with positive and negative responses fell into this category.
In some cases, these responses contained some elaboration, such as items OP liked or disliked.
% items they already knew, or items they didn't know.

To sum, these were the types of interactions we observed in \sys.
Current recommender systems can incorporate these characteristics to present better recommendations.
% , either through better modeling of the attributes inside the algorithm or through enabling customized filtration of the results.
We elaborate on the implications of these results in the Discussion section.

% \begin{figure}[t!]
%     \centering
%     \begin{subfigure}[t]{0.24\textwidth}
%         \centering
%         \includegraphics[width=0.95\textwidth]{figures/genre_1.pdf}
%         \caption{All music genres from everynoise.com}
%     \end{subfigure}%
%     ~
%     \begin{subfigure}[t]{0.24\textwidth}
%         \centering
%         \includegraphics[width=0.95\textwidth]{figures/genre_2.pdf}
%         \caption{Genres of music mentioned in \sys{} posts}
%     \end{subfigure}
%     ~
%     \begin{subfigure}[t]{0.24\textwidth}
%         \centering
%         \includegraphics[width=0.95\textwidth]{figures/genre_3.pdf}
%         \caption{Genres of music in four popular playlists on Spotify}
%     \end{subfigure}
%     \begin{subfigure}[t]{0.26\textwidth}
%         \centering
%         \includegraphics[width=1\textwidth]{figures/pair_plot.pdf}
%         \caption{Pair plot of Spotify playlists and \sys ~genres}
%     \end{subfigure}
%     % \vspace{-8pt}
%     \caption{Differences in presence of music genres.}
%     \label{fig:genre-comp}
% \end{figure}

\ssubsubsection{Discussion}
Sometimes, user interactions led to long conversations. 
We quantitatively identified long interaction chains.
We used heuristics of at least 3 back-and-forths; that is, chains of length 6.
This cutoff led us to identify 44 such conversations within our data set.
These discussions were not always between the OP and another commenter---some were between two or more other commenters.
These conversations fell into several themes.

\begin{itemize}
    \item Discussing a Particular Artist (4): Some conversations involved discussing a particular artist or band, their artifacts, and their evolution.
    \item Incremental Refinement of Recommendation (12): Some discussions between OP and a commenter revolved around incrementally refining suggestions as OP and the commenter went back and forth with restrictions. We further discuss refinement as a separate sub-code later.
    \item Sharing Playlists (2): When users realized they had similar tastes, they sometimes shared playlists with each other.
    \item Critiquing Item (Not) Matching (10): In many instances, two commenters discussed whether an item matched the criteria set by the OP.
    \item Discussing Requirement Homogeneity (4): Sometimes, commenters discussed whether items listed by the OP were homogenous, or whether the stated genre categorization made sense.
    \item Recommending Between Commenters (5): Some discussions led commenters to offer recommendations to each other.
\end{itemize}

\section{Discussion}
Our analysis revealed a set of characteristics with 9 high-level codes and 37 sub-codes describing how users ask \sys for recommendations, how the ensuing interactions unfold, and how commenters respond with recommendations.
Below, we discuss the implication of our results with respect to how one may go about looking for recommendation considering both algorithmic and social approaches; and the challenges in operationalizing the characteristics found in \sys{} to algorithmic systems.
% ; and \textbf{! UNFINISHED SENTENCE !}  

\subsection{When Should One Seek Recommendation in \sys{}?}
Results from our RQ1 reveal a model of how users generally come to \sys{} for recommendation online and how others can utilize the information.
For example, we found that users tend to exhaust other approaches first.
These approaches include using both general and special purpose search systems, and using the exploration systems built into the system like genre wise music playlists on Spotify.
As revealed in the quotes, trying the existing built-in system features over and over might not necessarily result in finding the desired results eventually.
This phenomena also indicate the limitations of modern algorithmic recommender systems where users are stuck with certain recommendations, no matter what.
In other cases, we also found that it is easier to come to \sys{} when someone do not know what they are looking for.
The reason behind not knowing could be that users are new to a genre of media or that a media does not perfectly fit into existing genres.
In both cases, recommendation seeking from \sys{} users could be far better than going into a search rabbit hole online.
\addition{This is particularly important, since a prior work showed that effort is one of important dimension for evaluation of recommendation system use~\cite{hosey2019just}. }
Finally, even when someone found some recommendations from search engines, asking at \sys{} is not necessarily a waste, since they can come across new discoveries from an expert stranger on a niche topic.
\addition{Scholars found users may prefer such new discoveries coming from anyone, including from strangers~\cite{bhuiyan2022othertube}.}
Overall, while many come to the subreddit after exhausting other options, that is not necessarily what someone should do.

\subsection{How to Succeed in Seeking Recommendations from Online Strangers}
Our investigation from RQ2 and RQ3 reveal important considerations for succeeding in recommendation seeking from strangers, from ``how to formulate queries'' to ``how to interact and dissect the results''.
Our second research question shows that while primary interest in a recommendation seeking activity could be an artifact or an artist, supporting information such as the context of the recommendation seeker, production/distribution information, and other additional details can also be important.
Take the example of recommendation seeking based on streaming platforms someone subscribes to.
Such information can certainly help responders to better provide suggestions.
Another important example is the time commitment of the requester.
When someone has little time, recommending media requiring longer time could be ineffective.
Such consideration is hardly taken into account by modern recommender systems, as found in our list.
At the same time, some of our subcodes reveal unique features in \sys.
For example, formatting-wise users can use abbreviations in their texts or share images/collages when seeking recommendation.
Furthermore, justifying how the items are similar is another important consideration to succeed in recommendation seeking.
Results from our third research questions reveal what users should expect in responses and interactions.
Sometimes, they may receive normal affirmation of tastes, while other times they may need to respond to questions from \sys{} users when the query is not clear or specific enough.
Therefore, articulating queries with as much specificity possible can help reduce questions.
\addition{Finally, our results may apply beyond media recommendation we focused on, in this study.
Many of the characteristics from Table \ref{tab:rq1} can apply to other types of recommendation. Take the example of recipe recommendation. Here, many subcodes like \textit{Theme}, \textit{Tools}, and \textit{Time Commitment}  apply during recipe seeking. However, there are some charactersitics that may not apply as well.
For instance, \textit{Completeness} may not apply for recipe recommendation, since users may not seek any incomplete recipe.
}
While we can elaborate on each subcode found in our results, for the sake of brevity, we will limit our discussion on the topic here.

\subsection{Design Dimensions for Recommender System}
Our investigation into \sys{} also reveal some design dimensions for considerations for recommendation systems.
The primary difference between recommendation seeking in traditional search systems and \sys{} is what powers the system: algorithm vs human.
This axis is an extension of the taxonomy mentioned earlier~\cite{roy2022systematic}.
On this human-algorithm axis, most algorithmic systems may not fall at the end of the spectrum for a few reason.
For example, most algorithmic systems are powered by user-item similarity which in part takes into account of choices of other users, especially in collaborative filtering systems~\cite{roy2022systematic}.
Now, literature also defines a class of recommender systems called social recommender systems that use users' social relationship inside its algorithm to assign recommendation~\cite{tang2013social}.
Here, reader should note that this social recommender systems are different than the human-powered mechanism of \sys.
In between fully algorithmic and fully human-powered recommender system design axis, there is also a class of algorithms called human-in-the-loop recommender system that combines the benefit of both systems~\cite{ustalov2022improving}.
Though human-in-the-loop systems are scarcely found in the real world, it is another consideration for recommender system designers.
When designing recommender systems, choice from the human-algorithm axis can affect other parts of the system.
It happens primarily because human-algorithm axis can correlate with other axis such as recommendation presentation axes (organized vs unorganized, aggregate vs not aggregate, random vs fixed).
System designers would have to consider benefits trade-offs of having aggregation capacity of algorithmic systems against  niche recommendation finding capacity of human-powered systems.

A second important dimension for consideration revealed from our investigation is the use of query by example dimension.
Whereas in typical systems one may seek recommendation based on a single example and its correlation with all other items~\cite{Vall2017}, unlike those systems, \sys{} enables recommendation solicitation based on a set of examples based on their latent similarity and new items similar to the latent attribute.
This process can help \sys{} to evade the popularity bias issue in algorithmic recommender systems~\cite{Chen2014}.
As revealed in RQ1, even when search systems allow someone to search by multiple queries, they may not realize which latent aspects users see the similarities in.
Consequently, for instance, when user is seeking recommendation for a movie based location similarity, Google may only return plot-based similar items.
Traditional recommender systems can consider designing recommendation systems with multiple examples and consider devising new features based on the latent similarity aspects requesters mention, as found in our result.

A third important dimension for consideration in recommendation systems design is the interactivity.
We found the interactivity allow room for discussion which enables recommendation seeker and recommenders to align any mismatch in understand the queries.
While ChatGPT like conversation recommender systems may enable new recommender systems to be interactive, what it may lack is the rationalization capacity of a human.
Therefore, recommender system designers need to consider this interactive rationalization capacity aspect in their system, which concurs with some of the recent literature~\cite{Cai2019, Cai2020, Lyu2021, liang2023enabling}. 
Still, our result does inform how conversational recommender system may provide some human-like experience by simulating subcodes such as taste affirmation, and providing justification.

\subsection{Challenges in Operationalizing the Characteristic Codes}
We have found a set of characteristics for consideration in the recommendation system design. 
This also brings up questions about how well we can operationalize them inside a recommender system.
Lets consider the possibility of using it a search filter.
If we examine our list, it may appear that not all characteristics are equal when it comes to operationalizing.
For example, operationalizing completeness of an artifact or demography of an artist only require of a finite set of options.
Scene the recommendation seeker, on the contrary, can be very arbitrary description.
Matching a description to a suitable recommendations can also be hard for characteristics in various ways.
First, some description may refer to other information sources, as we have seen inside query specificity (rhythm starting at ``0.32 second'' of a song).
Without deriving the corresponding content, effective recommendation can not be produced.
Second, there could be some subjective difference between how various users perceive an each option.
Take the example of activity level of an artifact.
To some, releasing one song could make an artist active, while for someone else it requires releasing an album to be active.
Third, some of the characteristics could be out of the scope for some recommender systems.
For example, while Google may provide recommendation from other artifact types, IMDB can not recommend a movie.
After matching, there are challenges in result presentation too.
One particular problem here is providing confidence in the recommendation through justification.
Despite the progress in recommender systems, this area of research is still not caught up to provide sufficient justification to the user.
Therefore, one potential direction of research is to examine how to operationalize the characteristics in recommender systems.
One potential direction of our future research is further exploration of these recommendation system design dimensions and how the dimensions affect recommendation seeking process.

\subsection{Limitations}
The findings from our study have the following limitations. 
First, since we opted to examine the data within a recent period and with some sampling, these results could paint an incomplete picture.
Second, users who interact with \sys{} do not represent the whole user population.
Since we sampled posts that got replies for qualitative coding, our findings have survivors' bias. Studying unanswered posts may reveal other characteristics in recommendation-seeking activities.
Thus, there might be some bias in the data we collected---and by extension, in our results.
% Third, some of our methods, including use of dictionaries, regular expressions or Google searches add their own uncertainties.
% Thus, the corresponding result should not be considered within those parameters, instead of absolutes.
Despite these limitations, our results provide a basis for exploration into how to improve recommendation-seeking using online strangers.
One direction for our future work involves exploring whether the incorporation of these characteristics into recommender systems can contribute to better user satisfaction in recommendation seeking.

% a conversation on users about the interaction between commenters and OP, the interactions that we found were rarely a back and forth. This is not too surprising seeing as how r/ifyoulikeblank is focused on person-to-person recommendation, however, finding conversations over topics are different from online recommender systems. This is a benefit to person-to-person recommendation. 
% \subsection{Beyond Popularity Bias}
% \subsection{Testing ChatGPT}
% Pros:
% \begin{itemize}
%     \item Answers in the same format
%     \item Provides rationale for why it thinks items in the questions are similar
%     \item Provides rationale for why it recommended a particular item
%     \item can take genre (dark metal) or demography (female vocal) into consideration
%     \item Can refine through a conversation
% \end{itemize}

% Cons:
% \begin{itemize}
%     \item It doesn't take images as input
%     \item It doesn't provide generalized suggestions outside the format
%     \item It would provide wrong recommendation based on wrong understanding of what makes items similar when reason is not given
%     \item It would provide wrong recommendation based on not understanding the reason
% \end{itemize}
\section{Conclusion}
In this work, we investigated \sys{} to understand the characteristics of queries, responses and corresponding interactions during recommendation seeking.
Our analysis reveals high-level dimensions and various underlying characteristics for consideration in such a recommendation-seeking setting, such as \textit{Artifact} (e.g., the tools used to create it), \textit{Artist} (e.g., their activity level), or \textit{OP's Context} (e.g., the activity performed in a context).
The results of this study can inform recommendation system developers on how to adapt their systems considering such characteristics.

\begin{acks}
    This paper would not be possible without the valuable feedback from the anonymous reviewers. Bhuiyan was partly supported by the College of Engineering Critical Needs Graduate Scholarship at Virginia Tech. 
\end{acks}
\bibliographystyle{ACM-Reference-Format}
\bibliography{main}


% \newpage
\appendix

\received{January 2024}
\received[revised]{April 2024}
\received[accepted]{Mayy 2024}

\end{document}